\documentclass[12pt]{article}
\usepackage[english]{babel}
\usepackage{graphicx}

   \usepackage{graphics,color,pstricks,pst-plot,pst-node,pst-coil}
   \usepackage[cp1251]{inputenc}

\begin{document}

\begin{center}
{\large \bf Natural Limits of Electroweak Model as Contraction of its Gauge Group}
\end{center}

\begin{center}
N.A.~Gromov \\
Department of Mathematics, Komi Science Center UrD RAS, \\
Kommunisticheskaya st. 24, Syktyvkar 167982, Russia \\
E-mail: gromov@dm.komisc.ru
\end{center}

\begin{abstract}
The low and higher energy limits of the  Electroweak Model
are obtained from  first principles of gauge theory.
Both limits are given by  the same contraction of the gauge group, but for the different consistent rescalings of the field space. Mathematical contraction parameter in both cases is interpreted as  energy.
The  very weak neutrino-matter interactions is  explained by
zero tending contraction parameter, which depend on neutrino energy.
The second consistent rescaling  corresponds to the higher energy limit of the Electroweak Model.
At the infinite energy all particles lose masses,
electroweak interactions become long-range
and are mediated by the neutral currents. The limit model represents the development of the early Universe from the Big Bang up to the end of the first second.
\end{abstract}

\section{Introduction }

The modern theory of electroweak processes  is the Electroweak Model, which is in good agreement with experimental dates, including the  latest ones from LHC. This model is a gauge theory  based on the gauge group $ SU(2)\times U(1)$, which is the direct product of two simple groups.
The operation of group contraction \cite{IW-53}
transforms a simple or semisimple group to a nonsemisimple one.
In particular the special unitary group $ SU(2)$ is contracted to the group isomorphic to Euclid group $E(2)$ \cite{GM-90,GM-92}.
For better understanding of a complicated physical system it is useful to investigate its behavior 
for limiting values of its physical parameters.
In this paper we  discuss mostly at the level of classical gauge fields the modified Electroweak Model with the contracted gauge group $ SU(2;j)\times U(1)$.
It was shown \cite{Gr-2012-a}--\cite{Gr-12}
that at low energies the contraction parameter depends on the energy $s$ in center-of-mass  system,  
so the contracted  gauge group corresponds to the zero energy limit of the Electroweak Model.
The  very weak neutrinos-matter interactions and
the linear dependence of their cross-section  on   neutrino energy  both   are explained  from  first principles of the   Electroweak Model as contraction of its gauge group.
But for the same contraction of the gauge group there is another consistent rescaling of the representation space,  which lead to the infinite energy limit of the Electroweak Model.
In this paper we consider  both possibilities and discuss some particle properties in early Universe, where similar higher energies can exist.

\section{Standard Electroweak Model } \label{s2}

 We shall follow the books \cite{R-99}--\cite{PS-95} in description of standard Electroweak Model.
Its  Lagrangian    is the sum of boson, lepton and quark Lagrangians 
\begin{equation}
L=L_B + L_L + L_Q.
\label{eq1}
\end{equation}
Boson sector $L_B=L_A + L_{\phi}$ involve two parts:
the gauge field Lagrangian
 \begin{equation}
L_A=\frac{1}{8g^2}\mbox{Tr}(F_{\mu\nu})^2 -\frac{1}{4}(B_{\mu\nu})^2=   -\frac{1}{4}[(F_{\mu\nu}^1)^2 +(F_{\mu\nu}^2)^2+(F_{\mu\nu}^3)^2] -\frac{1}{4}(B_{\mu\nu})^2
\label{eq2}
\end{equation}
   and the matter field Lagrangian
\begin{equation}
  L_{\phi}= \frac{1}{2}(D_\mu \phi)^{\dagger}D_\mu \phi -
  \frac{\lambda }{4}\left(\phi^{\dagger}\phi- v^2\right)^2,
\label{eq3}
\end{equation}
where
$ \phi= \left(
\begin{array}{c}
	\phi_1 \\
	\phi_2
\end{array} \right) \in C_2$
are the matter fields.
The covariant derivatives 
are given by
 \begin{equation}
D_\mu\phi=\partial_\mu\phi -ig\left(\sum_{k=1}^{3}T_kA^k_\mu \right)\phi-ig'YB_\mu\phi,
\label{eq4}
\end{equation}
where $T_k=\frac{1}{2}\tau_k, k=1,2,3$ are generators of $SU(2)$,
 $Y=\frac{1}{2}{\bf 1}$ is generator of $U(1)$, $g$ and $g'$ are constants.
The gauge fields
 \begin{equation}
A_\mu (x)=-ig\sum_{k=1}^{3}T_kA^k_\mu (x),\quad B_\mu (x)=-ig'B_\mu (x)
\label{gen}
\end{equation}
 take their values in Lie algebras $su(2),$  $u(1)$ respectively,  and the stress tensors are as follows
$$
F_{\mu\nu}(x)={\cal F}_{\mu\nu}(x)+[A_\mu(x),A_\nu(x)],\quad  B_{\mu\nu}=\partial_{\mu}B_{\nu}-\partial_{\nu}B_{\mu}.
$$

To generate   mass  for the vector bosons
the special mechanism of spontaneous symmetry breaking
is used.
 One of  $L_B$ ground states
$$
  \phi^{vac}=\left(\begin{array}{c}
	0  \\
	v
\end{array} \right), \quad  A_\mu^k=B_\mu=0
$$
is taken as a vacuum state of the model, and small field excitations $v+\chi(x) $
with respect to this vacuum  are regarded.

The fermion sector  is represented by the lepton $L_L$ and quark $L_Q$ Lagrangians. The lepton Lagrangian is taken in the form
\begin{equation}
L_L=L_l^{\dagger}i\tilde{\tau}_{\mu}D_{\mu}L_l + e_r^{\dagger}i\tau_{\mu}D_{\mu}e_r -
h_e[e_r^{\dagger}(\phi^{\dagger}L_l) +(L_l^{\dagger}\phi)e_r],
\label{eq14}
\end{equation}
where
$
L_l= \left(
\begin{array}{c}
	\nu_l\\
	e_{l}
\end{array} \right)
$
is the $SU(2)$-doublet,  $e_r $ is the $SU(2)$-singlet, $h_e$ is constant,
$\tau_{0}=\tilde{\tau_0}={\bf 1},$ $\tilde{\tau_k}=-\tau_k $, $\tau_{\mu}$ are  Pauli matrices and
$e_r, e_l, \nu_l $ are  two component Lorentz spinors.
Last  terms with factor $h_e$
represent electron mass.
The covariant derivatives  are given by the formulas:
$$
D_\mu L_l=\partial_\mu L_l -i\frac{g}{\sqrt{2}}\left(W_{\mu}^{+}T_{+} + W_{\mu}^{-}T_{-} \right)L_l-
$$
$$
 -i\frac{g}{\cos \theta_w}Z_\mu\left( T_3 -Q\sin^2 \theta_w  \right)L_l -ieA_\mu Q L_l,
$$
\begin{equation}
D_{\mu}e_r = \partial_\mu e_r -ig'QA_\mu e_r \cos \theta_w +ig'QZ_\mu e_r \sin \theta_w,
\label{eq5-10}
\end{equation}
where
$T_{\pm}=T_1\pm iT_2 $,
 $Q =Y+T_3$   is the generator of electromagnetic subgroup $U(1)_{em}$,
 $Y=\frac{1}{2}{\bf 1}$ is the hypercharge,
$ e=gg'(g^2+g'^2)^{-\frac{1}{2}} \;$ is the electron charge and
$ \sin \theta_w=eg^{-1}.$
The new gauge fields
$$
{ Z_\mu =\frac{1}{\sqrt{g^2+g'^2}}\left( gA_\mu^3-g'B_\mu \right)}, \quad
 { A_\mu =\frac{1}{\sqrt{g^2+g'^2}}\left( g'A_\mu^3+gB_\mu \right)},
$$
\begin{equation} 
{W_\mu^{\pm}=\frac{1}{\sqrt{2}}\left(A_\mu^1\mp iA_\mu^2  \right)}
  \label{eq5-1}
\end{equation}
are introduced instead of 
(\ref{gen}).

The quark Lagrangian is given by
$$
L_Q=Q_l^{\dagger}i\tilde{\tau}_{\mu}D_{\mu}Q_l + 
u_r^{\dagger}i\tau_{\mu}D_{\mu}u_r +
d_r^{\dagger}i\tau_{\mu}D_{\mu}d_r -
$$
\begin{equation}
-h_d[d_r^{\dagger}(\phi^{\dagger}Q_l) +(Q_l^{\dagger}\phi)d_r]
-h_u[u_r^{\dagger}(\tilde{\phi}^{\dagger}Q_l) +(Q_l^{\dagger}\tilde{\phi})u_r],
\label{eq14-Q}
\end{equation}
where left quark fields form the $SU(2)$-doublet
$
Q_l= \left(
\begin{array}{c}
	u_l\\
	d_{l}
\end{array} \right),
$
 right quark fields $u_r, d_r $ are the $SU(2)$-singlets,
$\tilde{\phi}_i=\epsilon_{ik}\bar{\phi}_k, \epsilon_{00}=1, \epsilon_{ii}=-1$ is the conjugate  representation of  $SU(2)$ group and  $h_u, h_d$ are constants.
All fields  $u_l, d_l, u_r, d_r $  are two component Lorentz spinors.
Last four terms with factors $h_d$ and $h_u$ specify the $d$- and $u$-quark mass.
The covariant derivatives     are given by
$$
D_{\mu}Q_l=\left(\partial_{\mu}- ig\sum_{k=1}^{3}\frac{\tau_k}{2}A^k_\mu -ig'\frac{1}{6}B_\mu  \right)Q_l,
$$
\begin{equation}
D_{\mu}u_r=\left(\partial_{\mu}-ig'\frac{2}{3}B_\mu  \right) u_r,\quad
D_{\mu}d_r=\left(\partial_{\mu}+ig'\frac{1}{3}B_\mu  \right) d_r.
\label{eq14-3Q}
\end{equation}

From the viewpoint of electroweak interactions  all known leptons and quarks are divided on three   generations.
Next two lepton generation are introduced in a similar way to (\ref{eq14}). They are left $SU(2)$-doublets
\begin{equation}
 \left(
\begin{array}{c}
	\nu_\mu\\
	\mu
\end{array} \right)_l, \quad
\left(
\begin{array}{c}
	\nu_\tau\\
	\tau
\end{array} \right)_l, \quad Y=-\frac{1}{2}
\label{eq14-1d}
\end{equation}
and right $SU(2)$-singlets:
$ 
\mu_r,  \tau_r,  Y=-1.
$ 
In addition to $u$- and $d$-quarks of the first generation there are
$(c,s)$ and  $(t,b)$ quarks of the next generations, which left fields
 \begin{equation}
 \left(
\begin{array}{c}
	c_l\\
	s_l
\end{array} \right), \quad
\left(
\begin{array}{c}
	t_l\\
	b_l
\end{array} \right), \quad Y=\frac{1}{6},
\label{eq14-1Q}
\end{equation}
are described by the  $SU(2)$-doublets and the right fields are  $SU(2)$-singlets:
$
c_r,\, t_r, \; Y=\frac{2}{3}; \;
s_r,\, b_r, \; Y=-\frac{1}{3}.
$
Their Lagrangians are introduced in a similar way to (\ref{eq14-Q}).
Full lepton and quark  Lagrangians are obtained by the summation over all generations.
In what follows we shall regarded only first generations of leptons and quarks.

\section{Modified  Electroweak Model } \label{s3}

We consider a model where the contracted gauge group
$SU(2;j)\times U(1)$  acts  in the boson, lepton and quark sectors. The contracted group $SU(2;j)$ is obtained \cite{Gr-2010,Gr-2011}
by the consistent rescaling of the fundamental representation of $SU(2)$  and the  space $C_2$
$$
z'(j)=
\left(\begin{array}{c}
jz'_1 \\
z'_2
\end{array} \right)
=\left(\begin{array}{cc}
	\alpha & j\beta   \\
-j\bar{\beta}	 & \bar{\alpha}
\end{array} \right)
\left(\begin{array}{c}
jz_1 \\
z_2
\end{array} \right)
=u(j)z(j), \quad
$$
\begin{equation}
\det u(j)=|\alpha|^2+j^2|\beta|^2=1, \quad u(j)u^{\dagger}(j)=1
\label{eq5-21}
\end{equation}
in such a way that   the hermitian form
\begin{equation}
z^\dagger z(j)=j^2|z_1|^2+|z_2|^2
\label{eq5-31}
\end{equation}
remains  invariant,
when contraction parameter  tends to zero
$j \rightarrow 0$ or is equal to the nilpotent unit $j= \iota,$  $\iota^2=0.$
The actions of the unitary group $U(1)$ and the electromagnetic subgroup $U(1)_{em}$
in the    space $C_2(\iota)$ with the base $\left\{z_2\right\}$ and the fiber $\left\{z_1\right\}$ are given by the same matrices as on the space $C_2$.

The  space $C_2(j)$ of the fundamental representation of $SU(2;j)$ group can be obtained from $C_2$ by substituting $z_1$ by $jz_1$.
Substitution $z_1 \rightarrow jz_1$ induces another ones for Lie algebra generators
$T_1 \rightarrow jT_1,\; T_2 \rightarrow jT_2,\;T_3 \rightarrow T_3. $
As far as the gauge fields take their values in Lie algebra, we can substitute the gauge fields instead of transforming the generators, namely:
\begin{equation}
A_{\mu}^1 \rightarrow jA_{\mu}^1, \;\; A_{\mu}^2 \rightarrow jA_{\mu}^2,\; \;A_{\mu}^3 \rightarrow A_{\mu}^3, \;\;
B_{\mu} \rightarrow B_{\mu}.
\label{g14}
\end{equation}
Indeed, due to commutativity and associativity   of multiplication by $j$
$$
su(2;j)\ni \left\{   A_{\mu}^1(j T_1) + A_{\mu}^2(j T_2) + A_{\mu}^3T_3\right\}=
$$
\begin{equation}
= \left\{ (j A_{\mu}^1)T_1 + (j A_{\mu}^2)T_2 +
A_{\mu}^3T_3\right\}.
\label{g14-d}
\end{equation}
For the  gauge fields (\ref{eq5-1})  these substitutions are as follows:
\begin{equation}
W_{\mu}^{\pm} \rightarrow jW_{\mu}^{\pm}, \;\; Z_{\mu} \rightarrow Z_{\mu},\; \;A_{\mu} \rightarrow A_{\mu}.
\label{g15}
\end{equation}
The left lepton
$
L_l= \left(
\begin{array}{c}
	\nu_l\\
	e_{l}
\end{array} \right)\;
$
and quark
$
Q_l= \left(
\begin{array}{c}
	u_l\\
	d_{l}
\end{array} \right)
$
fields are  $SU(2)$-doublets, so their  components are transformed in the similar way as  components of the vector $z$, namely:
\begin{equation}
 	\nu_l \rightarrow j\nu_l, \quad e_{l} \rightarrow e_{l}, \quad
 	u_l \rightarrow ju_l, \quad d_{l} \rightarrow d_{l}.
\label{g15-1}
\end{equation}
The right lepton and quark fields  are  $SU(2)$-singlets and therefore are not changed.

After transformations (\ref{g15}), (\ref{g15-1}) and spontaneous symmetry breaking 
the boson Lagrangian (\ref{eq2}),(\ref{eq3}) can be represented
in the form \cite{Gr-2010}--\cite{Gr-2010-a}

$$
 L_B(j)=L_B^{(2)}(j) + L_B^{int}(j)=
$$
$$
=\frac{1}{2}\left(\partial_\mu\chi \right)^2 -\frac{1}{2}m_{\chi}^2\chi^2
- {\frac{1}{4}{\cal Z}_{\mu\nu}{\cal Z}_{\mu\nu}+\frac{1}{2}m_Z^2Z_\mu Z_\mu}-
$$
$$
-{\frac{1}{4}{\cal F}_{\mu\nu}{\cal F}_{\mu\nu}} 
+j^2\left\{-{\frac{1}{2}{\cal W}_{\mu\nu}^{+}{\cal W}_{\mu\nu}^{-}+m_W^2W_\mu^{+}W_\mu^{-} }\right\} +
$$
\begin{equation}
 +L_B^{int}(j) =L_{B,b} + j^2 L_{B,f},
\label{eq5-2}
\end{equation}
where as usual  second order terms describe the boson particles content of the model and higher order terms $L_B^{int}(j)$  are regarded as their  interactions.
So Lagrangian (\ref{eq5-2}) include  charged $W$-bosons  with identical mass  $m_W=\frac{1}{2}gv$,
massless photon  $A_\mu, $
 neutral $Z$-boson  with the mass  $m_Z=\frac{v}{2}\sqrt{g^2+g'^2}$
and  scalar Higgs boson  $ \chi,\; m_{\chi}=\sqrt{2\lambda}v$.

In the limit   $j\rightarrow 0$ Lagrangian  (\ref{eq5-2}) is split on two parts: Lagrangian  
of fields in the base
$$
L_{B,b}=
\frac{1}{2}\left(\partial_\mu\chi \right)^2 -\frac{1}{2}m_{\chi}^2\chi^2
- {\frac{1}{4}{\cal Z}_{\mu\nu}^2 + \frac{1}{2}m_Z^2\left(Z_\mu\right)^2 }-
$$
$$
 -\frac{1}{4}{\cal F}_{\mu\nu}^2
+\frac{g m_z}{2\cos \theta_W} \left(Z_{\mu}\right)^2 \chi -\lambda v \chi^3 +
$$
\begin{equation}
+\frac{g^2 }{8\cos^2\theta_W} \left(Z_{\mu}\right)^2 \chi^2 - \frac{\lambda}{4} \chi^4
\label{5g}
\end{equation}
and Lagrangian  
of fields in the fiber
$$
L_{B,f}= -{\frac{1}{2}{\cal W}_{\mu\nu}^{+}{\cal W}_{\mu\nu}^{-} + m_W^2W_\mu^{+}W_\mu^{-} } -
$$
$$
-2ig\left(W_\mu^{+}W_\nu^{-} - W_\mu^{-}W_\nu^{+}\right)
\left({\cal F}_{\mu\nu}\sin \theta_W + \right.
$$
$$
\left.+{\cal Z}_{\mu\nu}\cos \theta_W\right) 
-\frac{i}{2}e \left[A_{\mu}\left({\cal W}_{\mu\nu}^{+}W_\nu^{-} - {\cal W}_{\mu\nu}^{-}W_\nu^{+}\right) - \right.
$$
$$
\left. -A_{\nu}\left({\cal W}_{\mu\nu}^{+}W_\mu^{-} - {\cal W}_{\mu\nu}^{-}W_\mu^{+}\right) \right]
+gW_\mu^{+}W_\mu^{-}\chi -
$$
$$
-\frac{i}{2}g\cos \theta_W  \left[Z_{\mu}\left({\cal W}_{\mu\nu}^{+}W_\nu^{-} - {\cal W}_{\mu\nu}^{-}W_\nu^{+}\right) - \right.
$$
$$
\left. -Z_{\nu}\left({\cal W}_{\mu\nu}^{+}W_\mu^{-} - {\cal W}_{\mu\nu}^{-}W_\mu^{+}\right) \right] +
$$
$$
+ \frac{g^2}{4}\left(W_\mu^{+}W_\nu^{-} - W_\mu^{-}W_\nu^{+}\right)^2 +
\frac{g^2}{4}W_\mu^{+}W_\nu^{-}\chi^2 -
$$
$$
-\frac{e^2}{4} \left\{
\left[\left(W_{\mu}^{+}\right)^2 + \left(W_{\mu}^{-}\right)^2\right](A_{\nu})^2 - \right.
$$
$$
-2\left(W_\mu^{+}W_\nu^{+} + W_\mu^{-}W_\nu^{-} \right)A_{\mu}A_{\nu} +
$$
$$
\left.+\left[\left(W_{\nu}^{+}\right)^2 + \left(W_{\nu}^{-}\right)^2\right](A_{\mu})^2
\right\} -
$$
$$
-\frac{g^2}{4}\cos\theta_W \left\{
\left[\left(W_{\mu}^{+}\right)^2 + \left(W_{\mu}^{-}\right)^2\right](Z_{\nu})^2 - \right.
$$
$$
 -2\left(W_\mu^{+}W_\nu^{+} + W_\mu^{-}W_\nu^{-} \right)Z_{\mu}Z_{\nu} +
$$
$$
\left. +\left[\left(W_{\nu}^{+}\right)^2 + \left(W_{\nu}^{-}\right)^2\right](Z_{\mu})^2
\right\} -
$$
$$
-eg\cos\theta_W \biggl\{
W_\mu^{+}W_\mu^{-}A_{\nu}Z_{\nu} +
W_\nu^{+}W_\nu^{-}A_{\mu}Z_{\mu} -
$$
\begin{equation}
-\frac{1}{2}\left(W_\mu^{+}W_\nu^{-} + W_\nu^{+}W_\mu^{-} \right)\left(A_{\mu}Z_{\nu} + A_{\nu}Z_{\mu}\right)
\biggr\}.
\label{6g}
\end{equation}

The lepton  Lagrangian (\ref{eq14}) in terms of electron and neutrino fields takes the form 
$$
L_L(j)=e_l^{\dagger}i\tilde{\tau}_{\mu}\partial_{\mu}e_l +
e_r^{\dagger}i\tau_{\mu}\partial_{\mu}e_r
-m_e(e_r^{\dagger}e_l + e_l^{\dagger} e_r)+
$$
$$
+\frac{g\cos 2\theta_w}{2\cos \theta_w}e_l^{\dagger}\tilde{\tau}_{\mu}Z_{\mu}e_l
-ee_l^{\dagger}\tilde{\tau}_{\mu}A_{\mu}e_l-
$$
$$
-g'\cos \theta_w e_r^{\dagger}\tau_{\mu}A_{\mu}e_r +
 g'\sin \theta_w e_r^{\dagger}\tau_{\mu}Z_{\mu}e_r +
$$
$$
+j^2\left\{ \nu_l^{\dagger}i\tilde{\tau}_{\mu}\partial_{\mu}\nu_l
 +
 \frac{g}{2\cos \theta_w} \nu_l^{\dagger}\tilde{\tau}_{\mu}Z_{\mu}\nu_l+ \right.
$$
\begin{equation}
\left.
+\frac{g}{\sqrt{2}}\left[ \nu_l^{\dagger}\tilde{\tau}_{\mu}W_{\mu}^{+}e_l +
 e_l^{\dagger}\tilde{\tau}_{\mu}W_{\mu}^{-}\nu_l\right]\right\}
 =L_{L,b}+ j^2L_{L,f}.
\label{g15-4}
\end{equation}

The quark  Lagrangian (\ref{eq14-Q}) in terms of $u$- and $d$-quarks  fields can be written as
$$
L_Q(j)=d_{l}^{\dagger}i\tilde{\tau}_{\mu}\partial_{\mu}d_{l}
+d_r^{\dagger}i\tau_{\mu}\partial_{\mu}d_r
-m_d(d_r^{\dagger}d_{l} + d_{l}^{\dagger}d_r)-
$$
$$
-\frac{e}{3}d_{l}^{\dagger}\tilde{\tau}_{\mu}A_{\mu}d_{l}
-\frac{g}{\cos \theta_w}\left(\frac{1}{2}-
\frac{2}{3}\sin^2\theta_w\right) d_{l}^{\dagger}\tilde{\tau}_{\mu}Z_{\mu}d_{l}-
$$
$$
 -\frac{1}{3}g'\cos\theta_w d_r^{\dagger}\tau_{\mu}A_{\mu}d_r +
\frac{1}{3}g'\sin\theta_w d_r^{\dagger}\tau_{\mu}Z_{\mu}d_r -
$$
$$
+j^2\biggl\{ u_{l}^{\dagger}i\tilde{\tau}_{\mu}\partial_{\mu}u_{l}
+u_r^{\dagger}i\tau_{\mu}\partial_{\mu}u_r
-m_u(u_r^{\dagger}u_{l} + u_{l}^{\dagger}u_r) +
$$
$$
+\frac{g}{\cos \theta_w}\left(\frac{1}{2}-\frac{2}{3}\sin^2\theta_w\right) u_{l}^{\dagger}\tilde{\tau}_{\mu}Z_{\mu}u_{l} +
$$
$$
+\frac{2e}{3}u_{l}^{\dagger}\tilde{\tau}_{\mu}A_{\mu}u_{l}
+\frac{g}{\sqrt{2}}\left[ u_{l}^{\dagger}\tilde{\tau}_{\mu}W^{+}_{\mu}d_{l} +
d_{l}^{\dagger}\tilde{\tau}_{\mu}W^{-}_{\mu}u_{l}\right]  +
$$
$$
+\frac{2}{3}g'\cos\theta_w u_r^{\dagger}\tau_{\mu}A_{\mu}u_r -
\frac{2}{3}g'\sin\theta_w u_r^{\dagger}\tau_{\mu}Z_{\mu}u_r
\biggr\}=
$$
\begin{equation}
=L_{Q,b} + j^2 L_{Q,f},
\label{QL}
\end{equation}
where $m_e=h_ev/\sqrt{2}$ and $m_u=h_uv/\sqrt{2},\; m_d=h_dv/\sqrt{2} $ represent electron and quark masses.

The full Lagrangian of the modified model is given by the sum
$$
L(j)=L_B(j) + L_Q(j) + L_L(j)=
$$
$$
=L_{B,b}+ L_{L,b}+ L_{Q,b} + j^2\left\{L_{B,f}+ L_{L,f}+ L_{Q,f} \right\}=
$$
\begin{equation}
 =L_b + j^2L_{f},
\label{eq14-Full}
\end{equation}
where $L_b  $ is Lagrangian of the base fields and $L_{f} $ is Lagrangian of the fiber fields.

The boson Lagrangian $L_B(j)$ was discussed in \cite{Gr-2010,Gr-2011}
on the level of classical fields,
where it was shown that masses of all particles involved in the Electroweak Model remain the same under contraction $j^2\rightarrow 0$.
In this limit the contribution $j^2L_{f}$ of  neutrino, $W$-boson and $u$-quark fields as well as their interactions with other  fields to the Lagrangian (\ref{eq14-Full}) will be vanishingly small in comparison  with contribution $L_b$ of electron, $d$-quark and remaining boson fields.
So  Lagrangian (\ref{eq14-Full}) describes very weak interaction of neutrino fields with the matter.
On the other hand, contribution of the neutrino part $j^2L_{f}$ to the full Lagrangian is risen when the parameter $j^2$ is increased, that  again corresponds to the experimental facts. So contraction parameter can be phenomenologically connected with neutrino energy.

\section{Decompositions of  physical systems and group contractions}

The standard way of describing a physical system  in field theory  is its decomposition  on independent  more or less simple subsystems and then introduction of interactions between them. In Lagrangian formalism this  imply that some terms describe independent subsystems (free fields) and the rest terms correspond to interactions between fields.
When subsystems are not interacted with each other the composed system is a formal unification of subsystems and symmetry group of the whole system is direct product $G=G_1\times G_2$, where
$G_1$ and $G_2$ are symmetry groups of the subsystems.
The Electroweak Model gives an example of such approach. Indeed, there are free  boson, lepton, quark fields in Lagrangian and terms which describe interactions between these fields.

The operation of group contraction transforms a simple or semisimple group $G$ to a nonsemisimple one with the structure of a semidirect product $G=A(\times G_1$, where $A$ is Abelian and $G_1\subset G$ is untouched subgroup.
At the same time the space of fundamental representation of group $G$ is  split  under contraction in such a way that subgroup $G_1$ acts in the fiber.
Gauge theory with contracted gauge group
describe a physical system, which break up into  two subsystems $S_b$ and $S_f$.
One subsystem $S_b$ include all fields from the base
and other subsystem $S_f$ is built from fiber fields.
$S_b$ form the close system since according to semi-Riemannian geometry \cite{P-65,Gr-09} properties of the base do not depend on points of the fiber, what means on the physical language   that fields from the fiber do not interact with  fields from the base.
But on the contrary properties of the fiber
depend on points of the base, therefore the subsystem $S_b$
exert influence upon $S_f$. More precisely fields from the base are outer (or background) fields for subsystem $S_f$ and specify  outer conditions in every fiber.

In particular, the simple group $SU(2)$ is contracted to the nonsemisimple one  $SU(2;\iota)$, which is isomorphic to the Euclid group $E(2)=A_2(\times SO(1)$, where Abelian subgroup $A_2$ is  generated by translations \cite{Gr-2010,Gr-2011}.
The  fields space of the standard Electroweak Model is split after the contraction in such a way that  neutrino, $W$-boson and $u$-quark fields are in the fiber, whereas all other fields are in the base.

In order to avoid terminological misunderstanding let us stress that we regard locally trivial spliting, which is defined by the projection in the field space.
This spliting  is understood in the context of  semi-Riemannian geometry \cite{P-65,Gr-09}, where properties of the base do not depend on the fiber and has nothing to do with the principal fiber bundle.

The simple and best known example of  fiber space is the nonrelativistic space-time with one  dimensional base, which is interpreted as  time, and  three dimensional fiber, which is interpreted as proper space.
It is well known, that in nonrelativistic
physics the time is absolute and does not depend on the space coordinates, while the space properties can be changed in time. The simplest  demonstration of this fact is 
Galilei transformation
$t'=t,\; x'=x+vt.$
The space-time of the special relativity is transformed to the nonrelativistic space-time when dimensionfull 
parameter ---
velocity of light $c$ --- tends to the infinity and dimensionless  parameter
$\frac{v}{c}$ 
tends to zero.

\section{Weak neutrino-matter interactions and the physical interpretation of the parameter $j$}

To discover  
the connection of gauge group contraction with limiting case of the Electroweak Model and establish
the physical meaning of the contraction parameter
we need 
more fine consideration on the level of quantized fields. Namely,
we  discuss 
neutrino elastic scattering on electron and quarks. The corresponding diagrams for  the neutral and charged currents interactions are represented in Fig.~1 and Fig.~2.
%
\begin{figure}[ht]
\includegraphics[width=\linewidth]{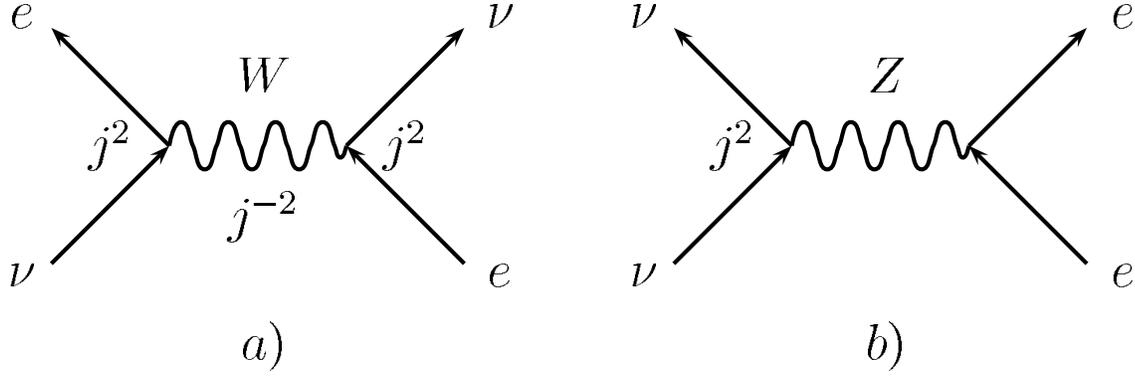}
\caption{Neutrino elastic scattering on electron }
\label{fig:1}
\end{figure}

\begin{figure}[ht]
\includegraphics[width=\linewidth]{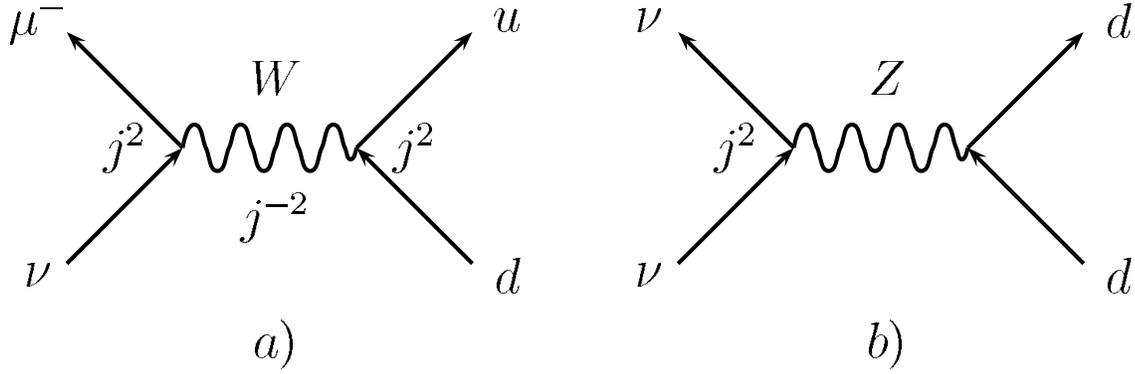}
\caption{Neutrino elastic scattering on quarks  }
\label{fig:2}
\end{figure}

%
Under substitutions (\ref{g15}),(\ref{g15-1}) both vertex of diagram in
  Fig.~1, a) are multiplied by $j^2$, as it follows from lepton Lagrangian (\ref{g15-4}). The propagator
of  virtual fields $W$ according to boson Lagrangian (\ref{eq5-2}) is multiplied by $j^{-2}$.
Indeed, propagator is inverse operator to  operator of free field, but the later for $W$-fields is multiplied by $j^{2}$.

So in total the probability amplitude for charged weak  current interactions is transformed as 
${\cal M}_W \rightarrow j^2{\cal M}_W.$  For diagram in Fig.~1, b) only one vertex is multiplied by $j^{2}$, whereas second vertex and propagator of $Z$ virtual field
 do not changed, so the corresponding amplitude for neutral weak  current interactions is transformed in a similar way
${\cal M}_Z \rightarrow j^2{\cal M}_Z.$ A cross-section is   proportionate to an squared amplitude, so neutrino-electron scattering cross-section 
is proportionate to $j^{4}$. For low energies $s \ll m_W^2$ this cross-section is as follows \cite{O-05}
\begin{equation}
\sigma_{\nu e} = G_F^2s f(\xi)=\frac{g^4}{m_W^4} \tilde{f}(\xi),
\label{cs-e}
\end{equation}
where
$G_F=10^{-5}\frac{1}{m_p^2}=1,17\cdot 10^{-5}\; GeV^{-2}$ is Fermi constant, $s$ is squared   
energy in central mass system, $\xi=\sin \theta_W$, $\tilde{f}(\xi)=f(\xi)/32$ is function of Weinberg angle.
The cross-section in the laboratory system for neutrino energy
$m_e \ll E_{\nu} \ll m_W$
is given by \cite{PDG-2010}
\begin{equation}
\sigma_{\nu e} = G_F^2 m_e E_{\nu} \tilde{g}(\xi).
\label{cs-e-11}
\end{equation}
On the other hand, taking into account that contraction parameter is dimensionless,
we can write down
\begin{equation}
\sigma_{\nu e} =j^4\sigma_0=(G_Fs)(G_F f(\xi))
\label{cs-e-1}
\end{equation}
and obtain
\begin{equation}
j^2(s)=\sqrt{G_Fs}\approx\frac{g\sqrt{s}}{m_W}.
\label{cs-e-2}
\end{equation}

Neutrino elastic scattering on  quarks due to neutral and charged currents
are pictured in Fig.~2. Cross-sections for neutrino-quarks scattering  are obtained in a similar way as for  the lepton case and are as follows \cite{O-05}
\begin{equation}
\sigma_{\nu}^W = G_F^2s \hat{f}(\xi)), \quad
\sigma_{\nu}^Z = G_F^2s\, h(\xi).
\label{cs-e-3}
\end{equation}
Nucleons are some composite construction of quarks, therefore some  form-factors are appeared in the expressions for neutrino-nucleons scattering cross-sections. The final expression
\begin{equation}
\sigma_{\nu n} = G_F^2s \hat{F}(\xi)
\label{cs-e-4}
\end{equation}
 coincide with (\ref{cs-e}), i.e. this cross-section is transformed as (\ref{cs-e-1})
with the contraction parameter (\ref{cs-e-2}).
At low energies scattering interactions   make the leading  contribution to
the total neutrino-matter  cross-section, therefore it has the same properties (\ref{cs-e-1}),(\ref{cs-e-2}) with respect to contraction of the gauge group.
So, the  very weak neutrinos-matter interactions and
the linear dependence 
of their cross-section  on   neutrino energy  both
   are explained by the   Electroweak Model  with the contracted gauge group.


\section{High-Energy Lagrangian of Electroweak Model  }

We shown in previous sections that contraction  $j \rightarrow 0$ of the gauge group (\ref{eq5-21}) of the Electroweak Model corresponds to its zero energy limit.
In this limit the first components of the lepton and quark doublets become infinite small in comparison with
their  second components. On the contrary, when energy  increase the first components of the  doublets become
greater then their  second ones.  In the infinite energy limit the second components of the lepton and quark doublets will be infinite small as compared with their  first components. To describe this limit we introduce new contraction parameter $\epsilon$ and {\bf new   consistent rescaling} of the group $SU(2)$  and the  space $C_2$
as follows
 $$
z'(\epsilon)=
\left(\begin{array}{c}
 z'_1 \\
\epsilon z'_2
\end{array} \right)
=\left(\begin{array}{cc}
	\alpha & \epsilon\beta   \\
-\epsilon\bar{\beta}	 & \bar{\alpha}
\end{array} \right)
\left(\begin{array}{c}
z_1 \\
\epsilon z_2
\end{array} \right)
=u(\epsilon)z(\epsilon),  
$$
\begin{equation}
\det u(\epsilon)=|\alpha|^2+\epsilon^2|\beta|^2=1, \quad u(\epsilon)u^{\dagger}(\epsilon)=1
\label{5}
\end{equation}
Both contracted groups $SU(2;j)$ (\ref{eq5-21}) and $SU(2;\epsilon)$ (\ref{5}) are the same and are isomorphic to Euclid group $E(2)$, but the space $C_2(\epsilon)$ is split  in the limit $\epsilon \rightarrow 0 $ on
 the one-dimension base $\left\{z_1\right\}$ and the one-dimension fiber $\left\{z_2\right\}$.
 From the mathematical point of view it is not important  first or second Cartesian axis  forms the base of fibering and in this sence  constructions (\ref{eq5-21}) and  (\ref{5}) are equivalent.
But the doublet components   are interpreted as a certain physical fields, therefore the fundamental representations (\ref{eq5-21}) and  (\ref{5}) of the same contracted  unitary group lead to the different limit cases of the Electroweak Model, namely, its zero energy and infinite energy limits.

In the second contraction scheme (\ref{5}) all gauge bosons are transformed according to the rules (\ref{g15})
with the natural substitution of $j$ by $\epsilon $. Instead of (\ref{g15-1}) the lepton and quark fields are transformed now as follows
\begin{equation}
 	 e_{l} \rightarrow \epsilon e_{l},  \quad  d_{l} \rightarrow \epsilon d_{l}, \quad
 	 \nu_l \rightarrow \nu_l, \quad  	u_l \rightarrow u_l.
\label{6}
\end{equation}
The next reason for inequality of the first and second doublet components is the special mechanism of spontaneous symmetry breaking, which is used to generate  mass of vector bosons and other elementary particles
of the model.
In this mechanism one of Lagrangian 
ground states
$
  \phi^{vac}=\left(\begin{array}{c}
	0  \\
	v
\end{array} \right) \;
$
is taken as vacuum of the model and then small field excitations
$ v+\chi(x) $ with respect to this vacuum are regarded.
So Higgs boson field $ \chi $ and constant $ v $ are multiplied by $\epsilon$.
As far as masses of all particles are proportionate to $ v $ we obtain the following transformation rule for the contraction (\ref{5})
\begin{equation}
 	\chi  \rightarrow \epsilon \chi,  \quad  v \rightarrow \epsilon v, \quad 
 m_p \rightarrow \epsilon m_p, \quad p={\chi}, W, Z, e, u, d.
\label{7}
\end{equation}

After  transformations (\ref{g15}), (\ref{6})--(\ref{7})
the boson Lagrangian of the Electroweak Model can be represented in the form
$$
 L_B(\epsilon)= - \frac{1}{4}{\cal Z}_{\mu\nu}^2 - \frac{1}{4}{\cal F}_{\mu\nu}^2 + \epsilon^2 L_{B,2}
 + \epsilon^3 gW_\mu^{+}W_\mu^{-}\chi  + \epsilon^4 L_{B,4},
$$
$$
 L_{B,4}= m_W^2W_\mu^{+}W_\mu^{-} -\frac{1}{2}m_{\chi}^2\chi^2 -\lambda v \chi^3  - \frac{\lambda}{4} \chi^4 +
 $$
 $$
 +\frac{g^2}{4}\left(W_\mu^{+}W_\nu^{-} - W_\mu^{-}W_\nu^{+}\right)^2 +
\frac{g^2}{4}W_\mu^{+}W_\nu^{-}\chi^2,
$$
$$
L_{B,2}= \frac{1}{2}\left(\partial_\mu\chi \right)^2
 + \frac{1}{2}m_Z^2\left(Z_\mu\right)^2 -\frac{1}{2}{\cal W}_{\mu\nu}^{+}{\cal W}_{\mu\nu}^{-} +
 $$
 $$
 +\frac{g m_z}{2\cos \theta_W} \left(Z_{\mu}\right)^2 \chi  + \frac{g^2 }{8\cos^2\theta_W} \left(Z_{\mu}\right)^2 \chi^2-
$$
$$
-2ig\left(W_\mu^{+}W_\nu^{-} - W_\mu^{-}W_\nu^{+}\right)
\Bigl( {\cal F}_{\mu\nu}\sin \theta_W + {\cal Z}_{\mu\nu}\cos \theta_W \Bigr)  -
$$
$$
-\frac{i}{2}e \left[A_{\mu}\left({\cal W}_{\mu\nu}^{+}W_\nu^{-} - {\cal W}_{\mu\nu}^{-}W_\nu^{+}\right) +
 \frac{i}{2}e A_{\nu}\left({\cal W}_{\mu\nu}^{+}W_\mu^{-} - {\cal W}_{\mu\nu}^{-}W_\mu^{+}\right) \right] -
$$
$$
-\frac{i}{2}g\cos \theta_W  \left[Z_{\mu}\left({\cal W}_{\mu\nu}^{+}W_\nu^{-} - {\cal W}_{\mu\nu}^{-}W_\nu^{+}\right) -
Z_{\nu}\left({\cal W}_{\mu\nu}^{+}W_\mu^{-} - {\cal W}_{\mu\nu}^{-}W_\mu^{+}\right) \right] -
$$
$$
-\frac{e^2}{4} \left\{
\left[\left(W_{\mu}^{+}\right)^2 + \left(W_{\mu}^{-}\right)^2\right](A_{\nu})^2 - \right.
$$
$$
\left.   -2\left(W_\mu^{+}W_\nu^{+} + W_\mu^{-}W_\nu^{-} \right)A_{\mu}A_{\nu} +
\left[\left(W_{\nu}^{+}\right)^2 + \left(W_{\nu}^{-}\right)^2\right](A_{\mu})^2
\right\} -
$$
$$
-\frac{g^2}{4}\cos\theta_W \left\{
\left[\left(W_{\mu}^{+}\right)^2 + \left(W_{\mu}^{-}\right)^2\right](Z_{\nu})^2 - \right.
$$
$$
\left. -2\left(W_\mu^{+}W_\nu^{+} + W_\mu^{-}W_\nu^{-} \right)Z_{\mu}Z_{\nu} +
\left[\left(W_{\nu}^{+}\right)^2 + \left(W_{\nu}^{-}\right)^2\right](Z_{\mu})^2
\right\} -
$$
$$
 -eg\cos\theta_W \left[
W_\mu^{+}W_\mu^{-}A_{\nu}Z_{\nu} +
W_\nu^{+}W_\nu^{-}A_{\mu}Z_{\mu} - \right.
$$
\begin{equation}
\left. -\frac{1}{2}\left(W_\mu^{+}W_\nu^{-} + W_\nu^{+}W_\mu^{-} \right)\left(A_{\mu}Z_{\nu} + A_{\nu}Z_{\mu}\right)
\right].
\label{8-2}
\end{equation}

The lepton  Lagrangian  in terms of electron and neutrino fields takes the form
$$
L_{L}(\epsilon)= L_{L,0} + \epsilon^2 L_{L,2} =
$$
$$
=\nu_l^{\dagger}i\tilde{\tau}_{\mu}\partial_{\mu}\nu_l +
e_r^{\dagger}i\tau_{\mu}\partial_{\mu}e_r  +g'\sin \theta_w e_r^{\dagger}\tau_{\mu}Z_{\mu}e_r -
$$
$$
- g'\cos \theta_w e_r^{\dagger}\tau_{\mu}A_{\mu}e_r
 + \frac{g}{2\cos \theta_w} \nu_l^{\dagger}\tilde{\tau}_{\mu}Z_{\mu}\nu_l +
 $$
 $$
 +\epsilon^2\Biggl\{ e_l^{\dagger}i\tilde{\tau}_{\mu}\partial_{\mu}e_l
-m_e(e_r^{\dagger}e_l + e_l^{\dagger} e_r)+
$$
$$
+ \frac{g\cos 2\theta_w}{2\cos \theta_w}e_l^{\dagger}\tilde{\tau}_{\mu}Z_{\mu}e_l
-ee_l^{\dagger}\tilde{\tau}_{\mu}A_{\mu}e_l +
$$
\begin{equation}%
\left. +\frac{g}{\sqrt{2}}\left( \nu_l^{\dagger}\tilde{\tau}_{\mu}W_{\mu}^{+}e_l +
 e_l^{\dagger}\tilde{\tau}_{\mu}W_{\mu}^{-}\nu_l\right)\right\}.
\label{9}
\end{equation}

The quark  Lagrangian  in terms of u- and d-quarks  fields
can be written as
$$
L_{Q}(\epsilon)= L_{Q,0} - \epsilon \, m_u(u_r^{\dagger}u_{l} + u_{l}^{\dagger}u_r) + \epsilon^2 L_{Q,2},
$$
$$
L_{Q,0}= d_r^{\dagger}i\tau_{\mu}\partial_{\mu}d_r
+u_{l}^{\dagger}i\tilde{\tau}_{\mu}\partial_{\mu}u_{l}
+u_r^{\dagger}i\tau_{\mu}\partial_{\mu}u_r-
$$
$$
-\frac{1}{3}g'\cos\theta_w d_r^{\dagger}\tau_{\mu}A_{\mu}d_r + \frac{1}{3}g'\sin\theta_w d_r^{\dagger}\tau_{\mu}Z_{\mu}d_r +
$$
$$
+\frac{2e}{3}u_{l}^{\dagger}\tilde{\tau}_{\mu}A_{\mu}u_{l}
+\frac{g}{\cos \theta_w}\left(\frac{1}{2}-\frac{2}{3}\sin^2\theta_w\right) u_{l}^{\dagger}\tilde{\tau}_{\mu}Z_{\mu}u_{l} +
$$
$$
+\frac{2}{3}g'\cos\theta_w u_r^{\dagger}\tau_{\mu}A_{\mu}u_r
-\frac{2}{3}g'\sin\theta_w u_r^{\dagger}\tau_{\mu}Z_{\mu}u_r,
$$


$$
L_{Q,2}=d_{l}^{\dagger}i\tilde{\tau}_{\mu}\partial_{\mu}d_{l}
- m_d(d_r^{\dagger}d_{l} + d_{l}^{\dagger}d_r)-
$$
$$
-\frac{e}{3}d_{l}^{\dagger}\tilde{\tau}_{\mu}A_{\mu}d_{l}
-\frac{g}{\cos \theta_w}\left(\frac{1}{2}-
\frac{2}{3}\sin^2\theta_w\right) d_{l}^{\dagger}\tilde{\tau}_{\mu}Z_{\mu}d_{l}+
$$
\begin{equation}
+\frac{g}{\sqrt{2}}\left[ u_{l}^{\dagger}\tilde{\tau}_{\mu}W^{+}_{\mu}d_{l} +
d_{l}^{\dagger}\tilde{\tau}_{\mu}W^{-}_{\mu}u_{l}\right].
\label{12}
\end{equation}

The complete  Lagrangian of the modified model is given by the sum
$
L(\epsilon)=L_B(\epsilon) + L_L(\epsilon) + L_Q(\epsilon)
$
and can be written in the form
\begin{equation}
L(\epsilon)=L_{\infty} + \epsilon L_1 + \epsilon^2 L_2 + \epsilon^3 L_3 + \epsilon^4 L_4.
\label{15}
\end{equation}
Unlike zero energy limit (\ref{eq14-Full}) 
the Electroweak Model demonstrate for $\epsilon \rightarrow 0$ five stages of behavior, which are  distinguished
by  powers of the contraction parameter.
 In the infinite energy   limit ($\epsilon =0$) Lagrangian is equal to
$$
L_{\infty}= - \frac{1}{4}{\cal Z}_{\mu\nu}^2 - \frac{1}{4}{\cal F}_{\mu\nu}^2 +
\nu_l^{\dagger}i\tilde{\tau}_{\mu}\partial_{\mu}\nu_l
+u_{l}^{\dagger}i\tilde{\tau}_{\mu}\partial_{\mu}u_{l}+
$$
\begin{equation}
+e_r^{\dagger}i\tau_{\mu}\partial_{\mu}e_r +
d_r^{\dagger}i\tau_{\mu}\partial_{\mu}d_r
+u_r^{\dagger}i\tau_{\mu}\partial_{\mu}u_r + L_{\infty}^{int}(A_{\mu},Z_{\mu}),
\label{15-dop}
\end{equation}
where
$$
L_{\infty}^{int}(A_{\mu},Z_{\mu})= \frac{g}{2\cos \theta_w} \nu_l^{\dagger}\tilde{\tau}_{\mu}Z_{\mu}\nu_l
+\frac{2e}{3}u_{l}^{\dagger}\tilde{\tau}_{\mu}A_{\mu}u_{l}+
$$
$$
+\frac{g}{\cos \theta_w}\left(\frac{1}{2}-\frac{2}{3}\sin^2\theta_w\right) u_{l}^{\dagger}\tilde{\tau}_{\mu}Z_{\mu}u_{l} +
$$
$$
+g'\sin \theta_w e_r^{\dagger}\tau_{\mu}Z_{\mu}e_r -
 g'\cos \theta_w e_r^{\dagger}\tau_{\mu}A_{\mu}e_r-
$$
$$
-\frac{1}{3}g'\cos\theta_w d_r^{\dagger}\tau_{\mu}A_{\mu}d_r + \frac{1}{3}g'\sin\theta_w d_r^{\dagger}\tau_{\mu}Z_{\mu}d_r +
$$
\begin{equation}
 + \frac{2}{3}g'\cos\theta_w u_r^{\dagger}\tau_{\mu}A_{\mu}u_r
-\frac{2}{3}g'\sin\theta_w u_r^{\dagger}\tau_{\mu}Z_{\mu}u_r.
\label{14}
\end{equation}

The limit model includes only {\bf massless particles}: 
photons $A_{\mu}$ and neutral  bosons $Z_{\mu}$, left quarks $u_{l}$ and neutrinos $\nu_l$, right electrons $e_r $ and  quarks $ u_r, d_r $.
The electroweak interactions become long-range because are mediated by the  massless neutral $Z$-bosons and photons. 

\begin{figure}[ht]
\begin{center}
\includegraphics [width=7.5cm, trim=0 0 0 0, clip]{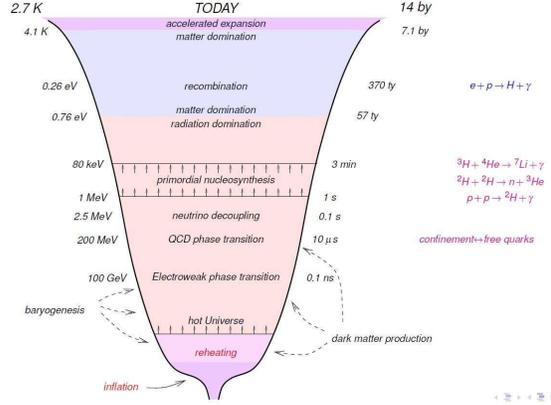}
\label{Schema}
\end{center}
\caption{History of the Universe ($1 eV=10^{4}K$) \cite{Gor-2013}}
\end{figure}

The infinite energies can exist only in the initial moment of  creation when the Universe is point-like \cite{L-1990, GR-2011}, as depicted in figure 3, and it is not clear what means  long-range electroweak interactions.
However more interesting is the Universe evolution  and
the limit Lagrangian $L_{\infty}$ can be considered as a good approximation near the  Big Bang
just as the nonrelativistic   mechanics is a good  approximation of the relativistic one at low velocities.
From the explicit form  of the interaction part 
$L_{\infty}^{int}(A_{\mu},Z_{\mu})$
it follows that there are no interactions between particles of different kind, for example neutrinos  interact only with each other by  neutral currents.
It looks like some stratification of the Electroweak Model with only one sort of particles in each stratum.

It is well known that to gain a
 better understanding of a physical system it is usefull to investigate its properties  for limiting values of  physical parameters.
It follows from (\ref{15}) that there are five stages in formation  of the Electroweak Model after the creation of the Universe which are distinguished by the powers of the contraction parameter $ \epsilon $. 
This offers an opportunity for construction of intermediate limit models. 
One can take the Lagrangian $L_{\infty}$ for the initial limit system, then add 
$L_1=-m_u(u_r^{\dagger}u_{l} + u_{l}^{\dagger}u_r) $
and obtain the second limit model with the Lagrangian
${\cal L}_1=L_{\infty} + \epsilon L_1 $. After that one can add 
$ L_2 $ and obtain the third limit model
${\cal L}_2=L_{\infty} + \epsilon L_1 + \epsilon^2 L_2 $. 
The last  limit model has the Lagrangian
${\cal L}_3=L_{\infty} + \epsilon L_1 + \epsilon^2 L_2 + \epsilon^3 L_3 $.
But it should noted  that among all limit models only $L_{\infty}$ is gauge model with the gauge group isomorphic to Euclid group $E(2)$.

Already at the level of classical gauge fields we can conclude that
the $u$-quark  first restores  its mass in the evolution of the Universe.
Indeed the mass term of $u$-quark in the Lagrangian (\ref{15})
is proportional to the first power $\epsilon L_1 $,
whereas the mass terms of $Z$-boson, electron and $d$-quark are multiplied by the second power of the contraction parameter 
\begin{equation}
\epsilon^2\, \left[\frac{1}{2}m_Z^2\left(Z_\mu\right)^2 + m_e(e_r^{\dagger}e_l + e_l^{\dagger} e_r)+
 m_d(d_r^{\dagger}d_{l} + d_{l}^{\dagger}d_r)\right]. 
\label{40}
\end{equation}
Higgs boson and charged $W$-boson, whose mass terms are multiplied by $\epsilon^4 $,  restore 
their masses after all other particles of the Electroweak Model.

\section{Conclusion} \label{s5}

We have investigated   the low and higher energy  limits of the Electroweak Model which are obtained from  first principles of gauge theory as contraction of its gauge group.
Above  limits are given by  the same contraction of the gauge group, but for the different consistent rescalings of the representation space. It was shown that mathematical contraction parameter in both cases is interpreted as typical energy.

The  very weak neutrino-matter interactions especially at low energies can be explained by this model already
at the level of classical (non-quantum) gauge fields.
The zero tending contraction parameter is connected with  neutrino energy  and reproduce the linear energy dependence of the neutrino-matter  cross-section.

The alternative rescaling of the gauge group and the field space corresponds to the infinite energy limit of the Electroweak Model, which goes in this limit through the five stages depending on the powers of the contraction parameters.
At the infinite energy all particles are massless and
electroweak interactions become long-range.
But the infinite energies can exist only in the Big Bang, i.e. in the initial moment of  creation when the Universe is point-like and it is not clear what means  long-range.
However more interesting is the Universe  development and the limit Lagrangian $L_{\infty}$ can be considered as a good approximation near the  Big Bang just as the nonrelativistic   mechanics is a good good approximation of the relativistic one at low velocities.
Particularly we can conclude that according to the Electroweak Model $u$-quark first  restores  its mass among other particles in the evolution of the Universe.



\begin{thebibliography}{99}
\bibitem{IW-53}
  In{\"o}n{\"u}~E. and Wigner~E.~P.  1953 
 On the Contraction of Groups and their Representations.
{\it Proc. Nat. Acad. Sci. U.S.A.}, 1953,  {\bf 39}  510--524. 
\bibitem{GM-90}
Gromov~N.~A. and  Man'ko~V.~I.  
The  Jordan-Schwinger representations  of  Cayley-Klein  groups. II.  The   unitary   groups.
{\it J. Math. Phys.}, 1990, {\bf 31},  1054--1059.
\bibitem{GM-92}
Gromov~N.~A. and  Man'ko~V.~I.  
Contractions of the irreducible representations  of  the  quantum  algebras  $ su_q(2) $ and $ so_q(3). $
{\it J. Math. Phys.}, 1992, {\bf 33},  1374--1378.
%
\bibitem{Gr-2012-a}
Gromov~N.~A. 
Contraction of Electroweak Model can Explain the Interactions Neutrinos with Matter.
{\it Phys. Part. Nucl.}, 2012, {\bf 43},  723--725.
%
\bibitem{Gr-2012}
Gromov~N.~A. 2012
Contraction of Electroweak Model and Neutrino.
{\it  Phys. Atom. Nucl.}, 2012, {\bf 75}, 1203--1209. 
%
\bibitem{Gr-2013}
Gromov~N.~A. 
Interpretation of Neutrino-Matter Interactions at Low Energies as Contraction  of   Gauge Group of Electroweak Model.
{\it Phys. Atom. Nucl.}, 2013, {\bf 76}, 1144--1148.
 %
\bibitem{Gr-12}
Gromov~N.~A.  {\it Contractions of Classical and Quantum Groups}. Fizmatlit, Moscow, 2012 [in Russian].
%
\bibitem{R-99}
Rubakov~V.~A. {\it Classical Gauge Fields}. Editorial URSS, Moscow, 1999 [in Russian].
%
%
\bibitem{O-05}
Okun'~L.~B. {\it Leptons and quarks}. Editorial URSS, Moscow,  2005 [in Russian].
%
\bibitem{PS-95}
Peskin~M.~E. and  Schroeder~D.~V. {\it An Introduction to Quantum Field Theory}. Addison-Wesley, 1995.
%
\bibitem{Gr-2010}
 Gromov~N.~A. 2010
Analog of Electroweak Model for Contracted Gauge Group.
{\it Phys. Atom. Nucl.}, 2010,  {\bf 73},  326--330.
\bibitem{Gr-2011}
Gromov~N.~A. 
 Limiting Case of Modified Electroweak Model for Contracted Gauge Group.
{\it Phys. Atom. Nucl.}, 2011,  {\bf 74}, 908--913.
%
 \bibitem{Gr-2010-a}
Gromov~N.~A. 
Higer and Low Energy Limits of Electroweak Model.
{\it  Proc. Komi SC UrB RAS}, 2014, {\bf 1(17)}, 5--9  
[in Russian].
 %
\bibitem{P-65}
Pimenov~R.~I. 
To the definition of  semi-Riemannian spaces,
{\it  Vestnik Leningrad Univ.}, 1965, {\bf 1}, 137--140  [in Russian].
%
%
\bibitem{Gr-09}
Gromov~N.~A. 
The R.I. Pimenov unified gravitation and electromagnetism field theory as semi-Riemannian geometry.
{\it Phys. At. Nucl.}, 2009, {\bf 72}, 794--800.
 %
%
\bibitem{PDG-2010}
Review of Particle Physics 2010,  http://pdg.lbl.gov, eq. (10.19)
\bibitem{L-1990}
Linde~L.~D.
{\it Particle Physics and Inflationary Cosmology}.
Nauka, Moscow, 1990 [in Russian].
%
\bibitem{GR-2011}
Gorbunov~D.~S. and Rubakov~V.~A.
{\it Introduction to the Theory of the Early Universe: Hot Big Bang Theory}.
 World Scientific, Singapure, 2011. 
\bibitem{Gor-2013}
Gorbunov~D.~S. 2013,
http://crimea.bitp.kiev.ua/reg/files/gorbunov.pdf.
\end{thebibliography}
\end{document}